\documentclass[12pt,a4paper]{article}
\pdfoutput=1

\usepackage{slashed,braket}
\usepackage{jheppub}
\usepackage{subcaption}

\usepackage{tikz}
\usetikzlibrary{arrows,shapes}
\usetikzlibrary{calc}				% For coordinates
\usetikzlibrary{decorations.pathreplacing}  % For curly braces
\usetikzlibrary{decorations.pathmorphing}	% For Feynman Diagrams
\usetikzlibrary{decorations.markings}

\tikzset{
    v/.style={decorate, decoration={snake}, draw,thick},
	provector/.style={decorate, decoration={snake,amplitude=2.5pt}, draw},
	antivector/.style={decorate, decoration={snake,amplitude=-2.5pt}, draw},
    f/.style={draw=black, postaction={decorate},
        decoration={markings,mark=at position .55 with {\arrow[very thick]{latex}}}},
    fb/.style={draw=black, postaction={decorate},
        decoration={markings,mark=at position .55 with {\arrow[very thick]{latex}}}},
    fnar/.style={draw=black,thick},
    g/.style={decorate, draw=black,
        decoration={coil,amplitude=3pt, segment length=4pt}},
    s/.style={dashed,draw=black, postaction={decorate},
        decoration={markings,mark=at position .55 with {\arrow[very thick]{latex}}}},
    sb/.style={dashed,draw=black, postaction={decorate},
        decoration={markings,mark=at position .55 with {\arrow[very thick]{latex}}}},
    snar/.style={dashed,draw=black}
}
\usepackage{tkz-euclide}
\usetkzobj{all}

\title{  
\vspace*{-2.3cm}  
\begin{flushright}  
\normalsize{\textnormal{  
SLAC-PUB-16318
  }}  
\end{flushright}  
\vspace{1.5cm}  
\begin{flushleft}\Huge  
\textbf{Sneutrino Higgs models explain lepton non-universality in CMS excesses
}\end{flushleft}\vspace*{1.0cm}   
}

\author[a]{Joshua Berger,}
\emailAdd{jberger@slac.stanford.edu} 
\affiliation[a]{SLAC National Accelerator Laboratory, 2575 Sand Hill Road, Menlo Park, CA 94025, USA}
\author[b]{Jeff Asaf Dror,}
\emailAdd{ajd268@cornell.edu} 
\author[b]{and Wee Hao Ng}
\emailAdd{wn68@cornell.edu} 
\affiliation[b]{Department of Physics, LEPP, Cornell University, Ithaca, NY 14853}

\abstract{Recent searches for first-generation leptoquarks and
  heavy right-handed $W_R$ bosons have seen excesses in final states
  with electrons and jets. A bizarre property of these excesses is
  that they appear to violate lepton universality.  With these results in mind, we study the phenomenology of
  supersymmetric models in which the Higgs arises as the sneutrino in
  an electron supermultiplet. Since the electron is singled out in this approach, one can naturally account for the
  lepton flavor structure of the excesses. In this work, we show that
  in such a framework, one can significantly alleviate the tension
  between the Standard Model and the data and yet evade current
  constraints from other searches. Lastly we point out that correlated
  excesses are expected to be seen in future multilepton searches.
}

 \arxivnumber{}

\begin{document} 
\maketitle 
\flushbottom

\section{Introduction}
The Standard Model (SM) of particle physics is among the most
successful models ever devised, yet it leaves open several puzzles
that should be resolved by a more complete description of nature.  A
well-motivated, broad class of models based on supersymmetry (SUSY)
has the potential to resolve one or more of the outstanding puzzles of
the SM, including the hierarchy problem, the nature of dark matter, the
mechanism of baryogenesis, and the running of gauge couplings to a
unified value.  From a phenomenological point of view, however, there
are several issues with models based on SUSY.  In particular, the naive implementation of 
natural R-parity conserving MSSM requires a light spectrum of
color-charged particles to which the experiments at the Large Hadron
Collider (LHC) should
have sensitivity, yet no hints of SUSY have been seen in the
``standard candle'' channels with Missing Transverse Energy~\cite{Craig2013}.
Furthermore, a Higgs boson with mass $125~{\rm GeV}$ is not
generically reconciled with a natural spectrum of superpartners~\cite{Hall:2011aa}.
Both of these tensions hint at the possibility that, if natural SUSY
describes our universe, then it may have an alternative structure.

The lack of observation at colliders has lead to the introduction of many variations of supersymmetry such as R-parity violating (RPV)~\cite{Nikolidakis:2007fc,Csaki:2011ge,Bhattacherjee:2013gr,Csaki:2013we,Franceschini:2013ne,Krnjaic:2013eta,Monteux:2013mna,DiLuzio:2013ysa,Csaki:2013jza} and $ R $-symmetric supersymmetry ~\cite{Kribs:2007ac,Benakli:2008pg,Amigo:2008rc,Benakli:2011vb,Benakli:2011kz,Heikinheimo:2011fk,Davies:2011jsm,Kumar:2011np,Fok:2012fb,Kalinowski:2011zzb,Frugiuele2011,Chakraborty:2013gea,Morita:2012kh,Frugiuele2013,Kalinowski:2011zz,Kalinowski:2011zzc,Frugiuele2011,Diessner:2014ksa,DeLopeAmigo:2011koa,Kribs:2013eua,Benakli:2012cy,Csaki:2013fla,% This last one has questionable relevance. (particular EWP test of R symmetric models)
Fok:2010vk}. Constraints on SUSY, even in the context of RPV models, are already quite stringent~\cite{Asano:2012gj,Han:2012cu,Berger:2013sir}. These constraints are somewhat less restrictive in models with R-symmetric models due to the requirement of Dirac gauginos~\cite{Fox2002}.  In particular, this prevents same-sign lepton signatures that would be smoking gun indicators of physics Beyond the SM (BSM). An additional intriguing feature of such models is that they allow for the Higgs field to be identified with the superpartner of a left-handed electron~\cite{Pomarol2012,Frugiuele2011,Frugiuele2012}\footnote{In general this can be any lepton, but as we will discuss in section~\ref{sec:intro}, the electron is the most natural choice.}. In this unique framework, traditional $ LLE^c$ and $LQD^c$ RPV effects are present but necessarily suppressed by the smallness of the Yukawa couplings. However, RPV effects appear due to a mixing between the electron doublet and the gauginos (such mixing has been previously used to put constraints of possible sneutrino VEVs~\cite{Brahm1990}). Since the electron is singled out as the Higgs partner, such models have non-standard lepton flavor structure leading in general to an abundance of electrons in the final state. Furthermore as we will show, the requirement of nearly massless neutrinos requires the introduction of an R-symmetry.

The CMS experiment has recently seen hints of potential BSM physics at the $\sim 2.5\sigma$ level in three separate searches that appear to single out the first generation of leptons.  Two of these analyses were optimized to look for pair production of leptoquarks.  In one case, the leptoquarks decay to an $eejj$ final state, while in the other they decay to an $e\nu jj$ final state~\cite{CMSLQ}.  Both showed excesses hinting at a roughly $650~{\rm GeV}$ leptoquark, at the $2.4 \sigma $ and $2.6\sigma$ levels respectively. The excesses are not consistent with the only decay modes of the leptoquarks being $ej$ and $\nu j$~\cite{CMSLQ,Bai2014}.  The third search was optimized for a $W_R$ decaying to an $eejj$ final state and saw a $2.8\sigma$ local excess for a resonance near $2.1~{\rm TeV}$~\cite{CMSWR}. However, the distributions of the excess do not appear to be consistent with those of a naive $ W_R $~\cite{CMSWR} (though see~\cite{Gluza2015} for a more general discussion on this possibility). Its important to note that the leptoquark searches did not see an excess in its high leptoquark mass bins. While not emphasized in earlier work, this puts serious limitations on new BSM signals attempting to explain the excess. No excesses were observed in the corresponding channels with muons~\cite{CMS-PAS-EXO-12-042,CMSWR}. 

Several models have been constructed in order to explain this excess.  Many of these models are supersymmetric in nature~\cite{Chun:2014jha,Allanach2014,Biswas2014,Allanach2014_2,Dhuria2015,Ovrut:2015uea} (see~\cite{Dobrescu:2015qna,Dobrescu:2015asa,Englert:2015oga,Allanach:2015ria,Evans2015,Gluza2015,Dhuria2015_2,Dhuria2015_3,Queiroz:2014pra,Dobrescu2014,Heikinheimo:2014tba,Bai2014,Mandal:2015vfa,Mandal:2015lca} for non-supersymmetric explanations).  The vast majority of them do not attempt to explain the puzzling lepton flavor structure of the observed excesses, but merely choose certain couplings to be larger then others.  Standard tools for suppressing flavor-violating processes such as minimal flavor violation (MFV)~\cite{D'Ambrosio:2002ex} cannot explain a different coupling for the first and second generations.  In MFV, such non-universal terms in the Lagrangian are suppressed by $m_\mu / m_\tau$.  Furthermore, due to the presence of a heavy resonance, these models often predict an excess in the searches for higher mass leptoquarks, which has not been observed in the data.

In this paper, we investigate the possibility that supersymmetric models with the Higgs as a sneutrino could explain the excesses seen by CMS.  The
lepton flavor structure is naturally obtained within the context of such models. The complex SUSY spectrum yields a rich variety of decay modes, suppressing the number of events seen in individual channels and allowing such models to evade many constraints.  Overall, this class of models provides a good fit for the current data, while making several new and testable predictions for the upcoming run of the LHC. The role of the leptoquarks in the model is played by a left-handed first-generation squark with R-parity violating decays, while the heavier $\sim 2~{\rm TeV}$ resonance is explained by gluino-squark production.  The masses that give the best fit are an up squark mass of $810~{\rm GeV}$ and a gluino mass of $1790~{\rm GeV}$.  In addition to accounting for the excesses observed by CMS, this model addresses the lack of an excess when the set of cuts is optimized for higher mass leptoquarks. The model considered in this paper addresses this potential issue by softening the ``leptoquark'' spectrum with additional jets, as proposed in~\cite{Dobrescu2014}.  % The presence of extra jets is a feature of many other models~\cite{} addressing the excesses and is an easy check of the compatibility of these models with the data. < -- JD: removed this line. I don't think "many" models actually considered extra jets

The remainder of this paper is structured as follows.  In Section 2, we review the minimal model with the Higgs as a sneutrino.  We determine a set of parameters of this model that provide a good fit to the current CMS data in Section 3.  We then conclude discussing current bounds on the model and provide additional predictions.

\section{Model with Higgs as a slepton}
\label{sec:intro}
\subsection{Overview}
To illustrate the main ideas behind the Higgs-as-slepton model~\cite{Pomarol2012}, we begin by attempting to construct a supersymmetric Standard Model that is more minimal than the MSSM. One can identify the SM Higgs doublet $H$ with a slepton doublet $\tilde{L}_a$, since they are both in the same gauge representation $(1,2)_{-1/2}$. The model then requires two fewer doublet chiral superfields than the MSSM. However, a major issue arises from the fact that the K\"{a}hler potential generates electroweak-scale Dirac masses between the partner leptons $L_a \equiv (\nu_a, l^-_a)$ and the Winos and Binos:
\begin{equation}
\mathcal{L} \supset 
-\frac{gv_H}{\sqrt{2}} l^-_a \tilde{W}^+ - \frac{gv_H}{2} \nu_a \tilde{W}^0 + \frac{g'v_H}{2} \nu_a \tilde{B} + \text{h.c.}
\end{equation}
This leads to neutrino masses that are too large. % In fact, similar considerations have been used to set bounds on the sneutrino VEVs in the MSSM \cite{Brahm1990}. NOTE: moved this to overview

One way around this difficulty is to first impose a $U(1)_R$ symmetry, with $R$-charge zero for the slepton doublet $\tilde{L}_a$ and $-1$ for the partner lepton doublet $L_a$. The $U(1)_R$ symmetry remains unbroken when $\tilde{L}_a$ acquires a VEV, and can still forbid Majorana masses for all $U(1)_R$-charged neutralinos.  By introducing adjoint chiral superfields $\Phi$ and SUSY-breaking Dirac gaugino masses, one of the neutralino mass eigenstates becomes massless. This massless neutralino is mainly comprised of $\nu_a$ and can be identified with the ``physical'' neutrino.

We now present the details of the model. Table~\ref{tab:model-summary} lists all the superfields and their gauge and $U(1)_R$ representations. With the CMS excesses in mind, we have chosen the first-generation leptons to partner the Higgs. This will give rise to experimental signatures specific to the electron, without the need to tweak any lepton couplings. $B$ and $L$ are arbitrary parameters that determine the $U(1)_R$ representations of the quark and the 2nd- and 3rd-generation lepton superfields.

\begin{table}
\centering
\renewcommand{\tabcolsep}{8pt}
\begin{tabular}[t]{| c | c | c|}
\hline
 & $SU(3)_C \times SU(2)_L \times U(1)_Y$ & $U(1)_R$\\
\hline
$H \equiv L _3$ & $(1,2)_{-1/2}$ & $0$\\
$E_3^c$ & $(1,1)_1$ & $2$\\
$L_{1,2}$ & $(1,2)_{-1/2}$ & $1-L$\\
$E_{1,2}^c$ & $(1,1)_1$ & $1+L$\\
$Q_{1,2,3}$ & $(3,2)_{1/6}$ & $1+B$\\
$U_{1,2,3}^c$ & $(\bar{3},1)_{-2/3}$ & $1-B$\\
$D_{1,2,3}^c$ & $(\bar{3},1)_{1/3}$ & $1-B$\\
$W^{a\alpha}$ & $(8,1)_0 + (1,3)_0 + (1,1)_0$ & $1$\\
$\Phi^{a}$ & $(8,1)_0 + (1,3)_0 + (1,1)_0$ & $0$\\
\hline
\end{tabular}
\caption{Superfields and their gauge and $U(1)_R$ representations in the Higgs-as-slepton model.}
\label{tab:model-summary}
\end{table}

The most general superpotential consistent with the symmetries (assuming $B \ne 1/3$ and $L \ne 1$) is
\begin{equation}
\mathcal{W} = \sum_{i,j = 1}^3 y_{d,ij} H Q_i D_j^c + \sum_{i = 1}^2 y_{e,i} H L_i E_i^c.
\end{equation}
We have chosen to work in the mass basis of the charged leptons. The superpotential does not generate up-type quark masses due to the absence of an up-type Higgs superfield $H_u$. The same is true for the electron mass, since the required term $H H E_3^c$ is identically zero. Both can be generated by SUSY-breaking K\"{a}hler terms of the form~\cite{Pomarol2012}
\begin{equation}
\int d^2\theta d^2 \bar{\theta} % x_{u,ij} 
\frac{X^\dagger}{M} \frac{H^\dagger Q_i U^c_j}{\Lambda}
\end{equation}
and
\begin{equation}
\int d^2\theta d^2 \bar{\theta} % x_e
\frac{X^\dagger X}{M^2} \frac{H D^\alpha H D_\alpha E^c_e}{\Lambda^2}
\end{equation}
that are suppressed by a $\Lambda$ cutoff scale. This also provides a natural explanation for the smallness of the electron mass, hence further motivating our decision to partner the first-generation leptons with the Higgs.

The $U(1)_R$ symmetry forbids mixing between left-handed and right-handed squarks, so the squark phenomenology differs from that of the MSSM~\cite{Frugiuele2013}. This also simplifies our subsequent analysis of squark production and decay since the squark mass eigenstates are then either left- or right-handed.

We note that the terms in the superpotential can also be interpreted as RPV terms of the form $L_3 Q_i D_j^c$ and $L_3 L_i E_j^c$. Therefore, experimental bounds on RPV coefficients~\cite{Barbier2005} can be applied to the superpotential Yukawas $y_{d,ij}$ and $y_{e,ij}$, which are in turn determined by the SM fermion masses and mixings. We find that these bounds are satisfied by the model for the choices of squark masses to be used in later sections.

While we assume the model described above in this work, our results are largely independent of the detailed mechanism giving the up-type quark and electron masses. Alternative models which introduce additional chiral superfields are also possible~\cite{Frugiuele2011,Frugiuele2012} and can also produce similar signatures. 
% Finally, we note that the Higgs-as-slepton model can be thought of as an even more minimal version of the More Minimal R-symmetric Supersymmetric Standard Model (MMRSSM) \cite{Frugiuele2011,Frugiuele2012}, a 2HDM model with two additional doublet chiral superfields $H_u$ and $R_d$ (for anomaly cancellation). The MMRSSM can also provide a viable explanation of the leptoquark anomaly through the same mechanism to be discussed in later sections, up to only minor modifications.

\subsection{Chargino and neutralino mass matrices and mixing}

The chargino and neutralino Dirac mass matrices are given by
\begin{equation}
\mathcal{M}_C = \bordermatrix{ & \tilde{W}^+ & \psi_{\tilde{W}}^+ & e_R^{c+}  \cr
\tilde{W}^- & 0 & M_{\tilde{W}} & 0 \cr
\psi_{\tilde{W}}^- & M_{\tilde{W}} & 0 & 0 \cr
e_L^- & \frac{g v_H}{\sqrt{2}} & 0 & 0}, \quad
\mathcal{M}_N = \bordermatrix{ & \tilde{B} & \tilde{W}^0  \cr
\psi_{\tilde{B}} & M_{\tilde{B}} & 0 \cr
\psi_{\tilde{W}}^0 & 0 & M_{\tilde{W}} \cr
\nu_e & -\frac{g' v_H}{2} & \frac{g v_H}{2}}
\end{equation}
We have neglected the masses from $\Lambda$-suppressed SUSY-breaking terms such as electron masses, since they are much smaller than the present terms and hence not expected to play an important role. To order $\epsilon \equiv g v_H /(2M_{\tilde{W}}) = m_W/M_{\tilde{W}}$, the chargino 4 component mass eigenstates are:

\begin{equation}
\chi^-_1 = \left( \begin{array}{c} -\sqrt{2}\epsilon \psi_{\tilde{W}}^- + e_L^- \\ e_R^- \end{array} \right), \quad
\chi^-_2 = \left( \begin{array}{c} \tilde{W}^- \\  \psi_{\tilde{W}}^{+ \, c} \end{array} \right), \quad
\chi^-_3 = \left( \begin{array}{c} \psi_{\tilde{W}}^- + \sqrt{2} \epsilon e_L^- \\ \tilde{W}^{+\, c} \end{array} \right)
\end{equation}
with mass eigenvalues $m_{\chi _1 ^- } = 0$ and $m _{ \chi _2 ^- } = m_{\chi _3 ^- } = M_{\tilde{W}}$. The mass eigenstates for the neutralinos are\footnote{We have assumed here that $\left|  M_{\tilde{W}}^2 - M_{\tilde{B}}^2 \right|  \gg m_W^2$. In the converse case where $\left|  M_{\tilde{W}}^2 - M_{\tilde{B}}^2 \right|  \ll m_W^2$, the actual heavy neutralino eigenstates are linear superpositions of $\chi_2^0$ and $\chi_3^0$ above, with mixings given by the Weinberg angle $\theta_W$. Nonetheless, this does not affect any of our subsequent results on the partial widths.}
:
\begin{equation}
\chi^0_1 = \left( \begin{array}{c} \frac{g'}{g} \frac{M_{\tilde{W}}}{M_{\tilde{B}}} \epsilon \psi_{\tilde{B}} - \epsilon \psi_{\tilde{W}}^0 + \nu_e \\ 0 \end{array} \right), \quad
\chi^0_2 = \left( \begin{array}{c} \psi_{\tilde{W}}^0 + \epsilon \nu_e \\ \tilde{W}^{0 \, c} \end{array} \right), \quad
\chi^0_3 = \left( \begin{array}{c} \psi_{\tilde{B}} - \frac{g'}{g} \frac{M_{\tilde{W}}}{M_{\tilde{B}}} \epsilon \nu_e \\  \tilde{B}^c \end{array} \right)
\end{equation}
with mass eigenvalues $m_{\chi _1 ^0 } = 0$, $m_{\chi _2 ^0 } = M_{\tilde{W}}$ and $m_{\chi _3 ^0 } = M_{\tilde{B}}$.

$\chi_1^-$ can be identified with the physical electron, and $\chi_1^0$ with the ``physical'' electron neutrino, before PMNS mixing.  We note that the gauge couplings of the physical gauginos and first-generation leptons to $W^\pm$ and $Z$ are affected by the $O(\epsilon)$ mixing. One consequence is that the $eeZ$ coupling is modified, hence violating lepton flavor universality. This allows us to place a lower bound of $\sim 2 \, \text{TeV}$ on the Dirac chargino mass $M_{\tilde{W}}$~\cite{Pomarol2012}. Another consequence is that the modified gauge couplings mix the physical gauginos and leptons, thus providing a channel for the gauginos to decay completely to SM particles, e.g. $\chi_2^0 \rightarrow \chi_1^- \, W^+$. Should the squarks be lighter than the gauginos, which we assume in the rest of this work, virtual cascades such as $\tilde{d}_L \rightarrow d \, \overline{\chi_2^0}^* \rightarrow d \, \overline{\chi_1^-} \, W^-$ may also become important decay channels for the first-generation squarks, as we will see below.

\subsection{First-generation left-handed squark decays}

\begin{figure}[t]
\begin{center}
\begin{tikzpicture} 
\draw[snar] (0.5,0) -- (2,0) node[pos=0,left] {$\tilde{q} $}; 
\draw[fnar] (2,0) -- (4,1.5) node[right] {$q' $}; 
\draw[fnar] (2,0) -- (3,-1.25) node[] at (1.25,-0.75) {$  \chi  ^{ 0 }_2 ,\chi  ^{ 0 }_3, \chi  ^{ -  }_2$}; 
\draw[v] (2,0) -- (3,-1.25); 
\draw[v] (3,-1.25) -- (4,-0.5) node[right] {$W, Z ,h $}; 
\draw[fnar] (3,-1.25) -- (4,-2.5) node[right] {$\chi  _1 ^- ,\chi  _1 ^0 $}; 
\end{tikzpicture}
\end{center}
\caption{Mixing-induced decay channels in which a supersymmetric particle $\tilde{q}_L$ decays completely to SM particles.} \label{fig:squark-decays}
\end{figure}
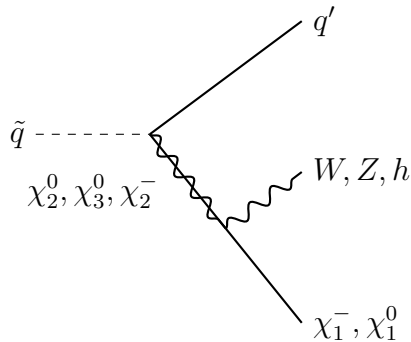

\begin{figure}[t]
\centering
\includegraphics[width=0.6\linewidth]{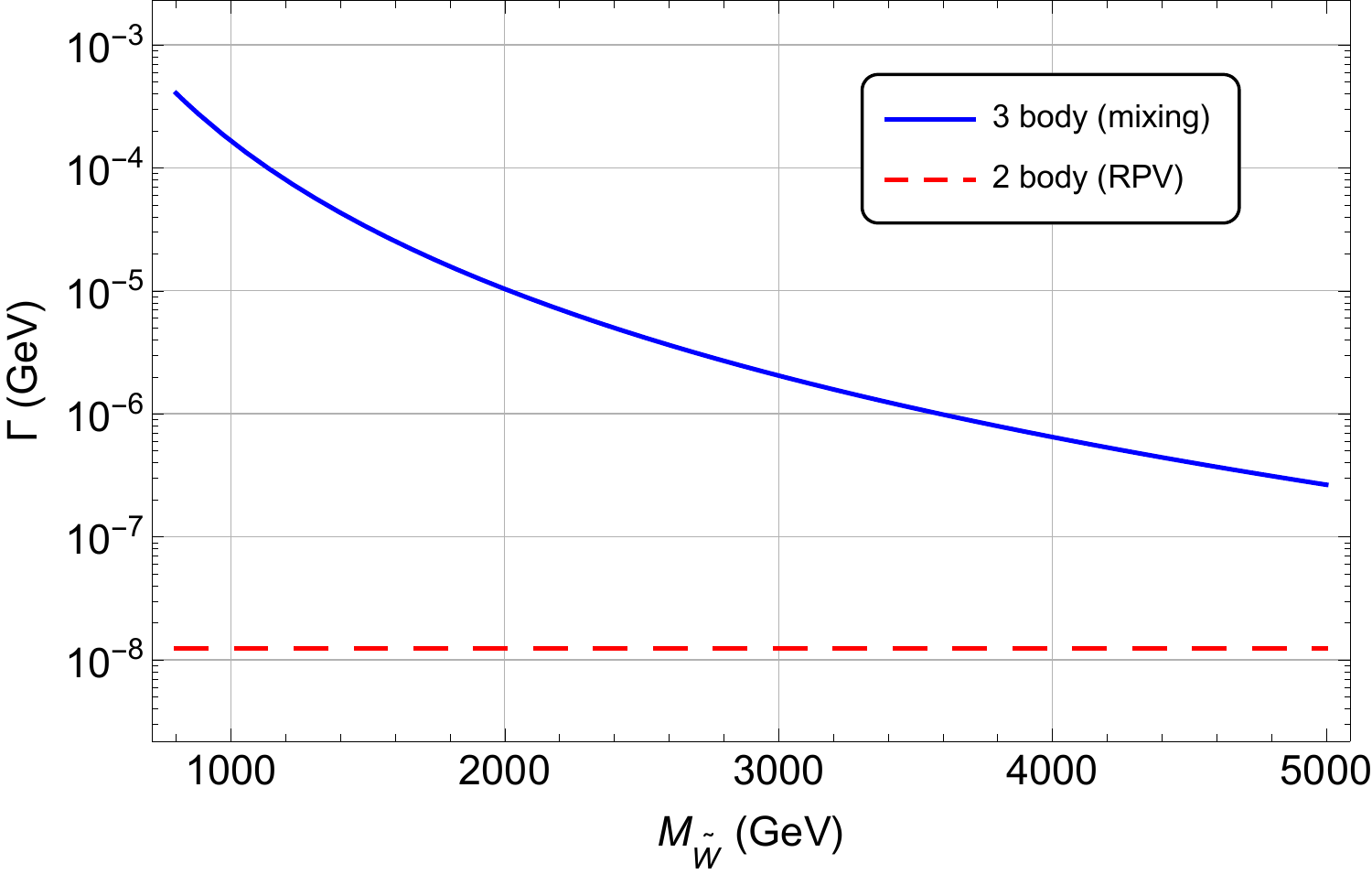}
\caption{Partial widths of $\tilde{d}_L$ for mixing-induced and standard RPV decay channels, assuming $m_{\tilde{d}}=810 \, \text{GeV}$ and $M_{\tilde{B}} = M_{\tilde{W}}$. The mixing-induced channel dominates over the range of $M_{\tilde{W}}$ considered.} \label{fig:comparing-decays}
\end{figure}

In MSSM with RPV, supersymmetric particles can decay completely to SM particles through channels generated by RPV superpotential and soft SUSY-breaking terms. While this is also true for the Higgs-as-slepton model, there are new decay channels due to the mixing of physical gauginos and leptons by the modified gauge couplings. A typical diagram for the new channel is shown in Fig.~\ref{fig:squark-decays}. The new channels are especially important for first-generation squarks compared to the standard RPV channels, due to the smallness of the Yukawas in the latter~\cite{Pomarol2012}. The approximate partial widths of these channels for first-generation LH squarks are shown in Table~\ref{tab:partial-widths}. Fig.~\ref{fig:comparing-decays} compares the partial widths of the mixing-induced and standard RPV channels for $\tilde{d}_L$ decay, from which we see that the former is dominant except for very large values of $M_{\tilde{W}}$. 

\begin{table}
\renewcommand{\arraystretch}{1.3}
$$
\begin{array}{|l|l|}
\hline 
\text{Decay channel} & \text{Partial width} \Gamma, /(\frac{1}{6144 \pi^3})\\[3pt]
\hline
\tilde{u}_L \rightarrow d \, \overline{\chi_1^-} \, h^0 & m_{\tilde{u}}^5 g^4/M_{\tilde{W}}^4 \times 1/2\\
\tilde{u}_L \rightarrow d \, \overline{\chi_1^-} \, Z & m_{\tilde{u}}^5 g^4/M_{\tilde{W}}^4 \times 1/2\\
\tilde{d}_L \rightarrow u \, \overline{\chi_1^0} \, W^- & m_{\tilde{d}}^5 g^4/M_{\tilde{W}}^4\\[3pt]
\hline
\tilde{u}_L \rightarrow u \, \overline{\chi_1^-} \, W^- & m_{\tilde{u}}^5 \left[ g^{\prime2} Y_Q /M_{\tilde{B}}^2 + g^2/(2 M_{\tilde{W}}^2) \right]^2 \\ 
\tilde{d}_L \rightarrow d \, \overline{\chi_1^0} \, h^0 & m_{\tilde{d}}^5 \left[ g^{\prime2} Y_Q /M_{\tilde{B}}^2 + g^2/(2 M_{\tilde{W}}^2) \right]^2 \times 1/2\\
\tilde{d}_L \rightarrow d \, \overline{\chi_1^0} \, Z & m_{\tilde{d}}^5 \left[ g^{\prime2} Y_Q /M_{\tilde{B}}^2 + g^2/(2 M_{\tilde{W}}^2) \right]^2 \times 1/2\\[3pt]
\hline
\tilde{u}_L \rightarrow u \, \overline{\chi_1^0} \, h^0 & m_{\tilde{u}}^5 \left[ g^{\prime2} Y_Q /M_{\tilde{B}}^2 - g^2/(2 M_{\tilde{W}}^2) \right]^2 \times 1/2\\
\tilde{u}_L \rightarrow u \, \overline{\chi_1^0} \, Z & m_{\tilde{u}}^5 \left[ g^{\prime2} Y_Q /M_{\tilde{B}}^2 - g^2/(2 M_{\tilde{W}}^2) \right]^2 \times 1/2\\
\tilde{d}_L \rightarrow d \, \overline{\chi_1^-} \, W^- & m_{\tilde{d}}^5 \left[ g^{\prime2} Y_Q /M_{\tilde{B}}^2 - g^2/(2 M_{\tilde{W}}^2) \right]^2\\[3pt]
\hline
\end{array}
$$
\caption{Partial widths for the mixing-induced decay channels. Here $\chi_1^-$ and $\chi_1^0$ refer to the physical electron and electron neutrino. $Y_Q$ is the hypercharge of the LH quark doublet. The decay channels have been arranged such that the approximate isospin symmetry from the Goldstone boson equivalence theorem is obvious.}
\label{tab:partial-widths}
\end{table}

Supersymmetric particles (and the Higgs) can also decay into SM particles + the gravitino, which is the lightest supersymmetric particle (LSP) in the model. The decay occurs via goldstino interaction terms fixed by supersymmetry, with partial widths that typically scale as $m_{\text{sp}}^5/(m_{3/2} M_{\text{Pl}})^2$, where $m_{\text{sp}}$ is the sparticle mass, $m_{3/2}$ the gravitino mass and $M_{\text{Pl}}$ the Planck scale~\cite{Pomarol2012}. However, as long as the gravitino mass is not too small ($m_{3/2} \gg 1 \, \text{eV}$), these decays are expected to be sub-dominant and can hence be neglected. For the rest of this work, we assume all first-generation squarks to decay via the mixing-induced decay channels.

%%% Local Variables:
%%% mode: latex
%%% TeX-master: "bobte"
%%% End:

\section{Simulation and Results}
In this section, we estimate the contribution of the above model to the CMS leptoquark and $W_R$ searches. The spectrum and production channels of interest are depicted in figures \ref{fig:spectrum} and \ref{fig:production}.
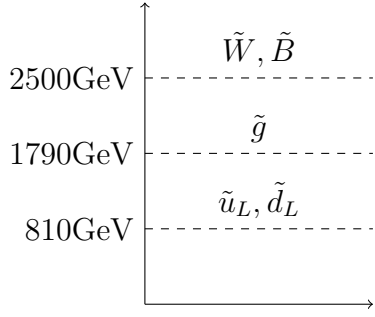
\begin{figure} 
\begin{center} \begin{tikzpicture} 
    \draw[->] (0,0) -- (0,4);
    \draw[->] (0,0) -- (3,0);
    \draw[dashed] (0,1) node[left] {$ 810 \mbox{GeV}$} -- (3,) node[midway,above] {$\tilde{u} _L , \tilde{d} _L $};
    % \draw[dashed] (0,2*5/9) node[left] {$ \sim 650 \mbox{TeV}$} -- (3,2*5/9) node[right] {$ \tilde{b} _R , \tilde{t} _R $};
    \draw[dashed] (0,2) node[left] {$1790 \mbox{GeV}$} -- (3,2) node[midway, above] {$ \tilde{g} $};
    % \node[above] at (1.5,2) ;
    \draw[dashed] (0,3) node[left] {$2500 \mbox{GeV}$} -- (3,3) node[midway, above] {$ \tilde{W}, \tilde{B}  $};
  \end{tikzpicture}
\end{center}
\caption{The spectrum of our benchmark point. All other fields are decoupled.}
\label{fig:spectrum}
\end{figure}
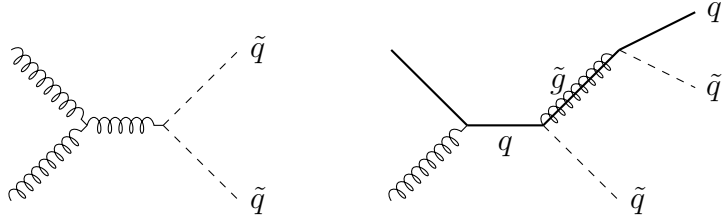
\begin{figure} 
\begin{center} \begin{tikzpicture} 
\draw[g] (0,0) -- (1,1) node[pos=0,left] {$$}; 
\draw[g] (0,2) -- (1,1) node[pos=0,left] {$$}; 
\draw[g] (1,1) -- (2,1) node[pos=0,left] {$$};
\draw[snar] (2,1) -- (3,2) node[right] {$\tilde{q} $};
\draw[snar] (2,1) -- (3,0) node[right] {$\tilde{q} $};

\begin{scope}[shift={(5,0)}]
\draw[g] (0,0) -- (1,1) node[pos=0,left] {$$}; 
\draw[fnar] (0,2) -- (1,1) node[pos=0,left] {$$}; 
\draw[fnar] (1,1) -- (2,1) node[midway,below] {$q$};
\draw[g] (2,1) -- (3,2) ; \node[] at (2.2,1.6) {$\tilde{g}$};
\draw[fnar] (2,1) -- (3,2) node[pos=0,left] {$ $};
\draw[snar] (2,1) -- (3,0) node[right] {$\tilde{q} $};
\draw[snar] (3,2) -- (4,1.5) node[right] {$\tilde{q} $};
\draw[fnar] (3,2) -- (4,2.5) node[right] {$q $};
\end{scope}
\end{tikzpicture}
\end{center}
\caption{Sample production mechanisms for disquark and single gluino production channels. Squarks decay through the $3$ body decay shown in figure~\ref{fig:squark-decays}.}
\label{fig:production}
\end{figure}

The model predictions are calculated at tree level using Madgraph~\cite{Madgraph}, Pythia 6.4~\cite{Pythia} for showering and hadronization, and PGS~\cite{PGS} for detector simulation. The model files were created using Feynrules~\cite{Feynrules}. To estimate the next-to-leading order (NLO effects we scaled the cross-sections by their corresponding K-factors calculated using Prospino 2.1~\cite{Prospino}. While Prospino was designed for the MSSM we do not expect significant deviations in the calculations of K-factors.

\begin{figure}
\centering
  \includegraphics[width=8cm]{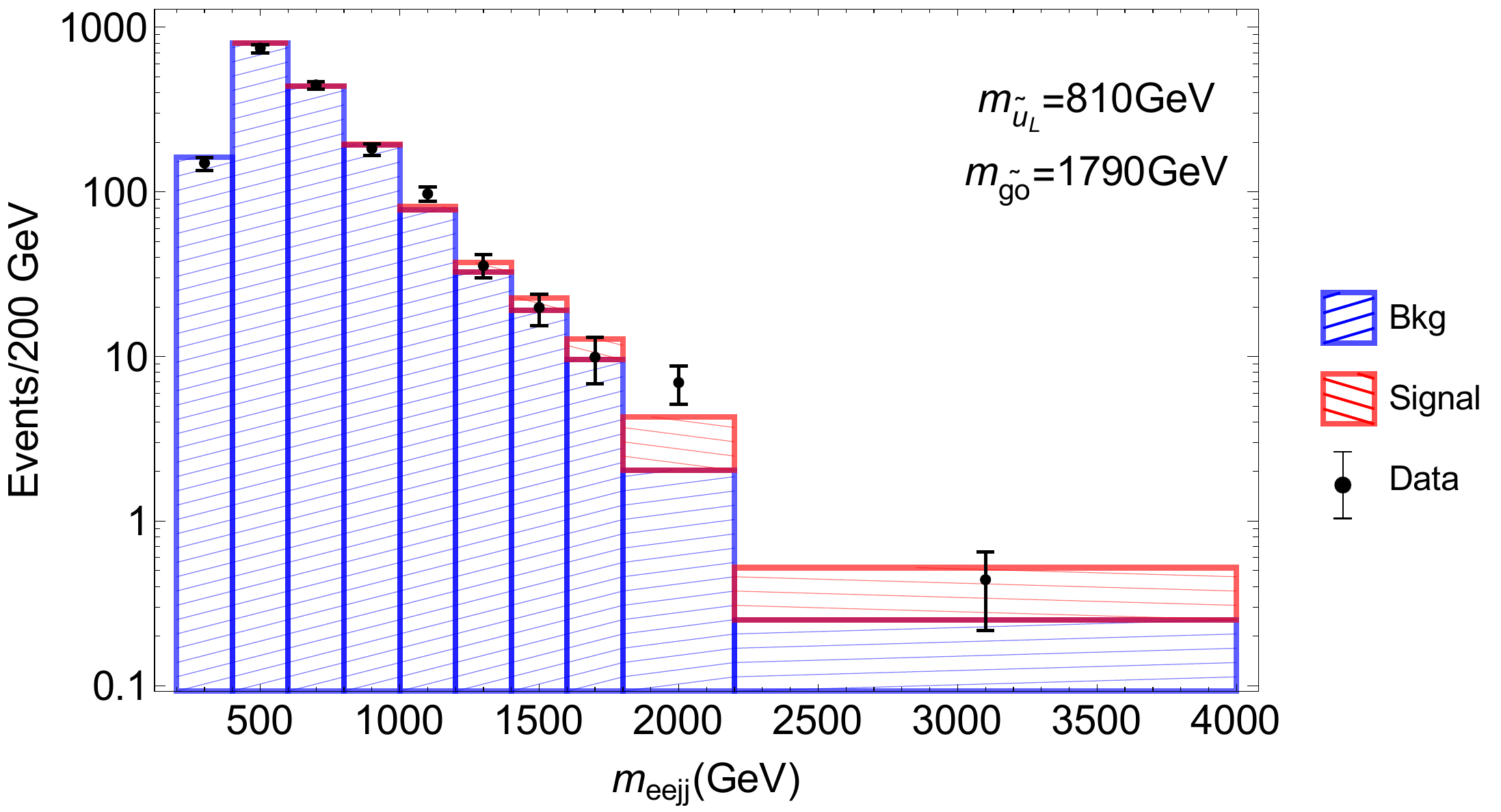}
 \caption{The $ m_{e e jj} $ prediction for our model after applying cuts used in the $ W _R $ search. The background and relevant cuts were taken from~\cite{CMSWR}.}
\label{fig:WR}
\end{figure}

The $ W _R $ search distribution is shown in figure \ref{fig:WR}.
We reproduce the invariant mass distribution of the two leading electrons and two leading jets. We also applied all the relevant cuts detailed by CMS in Ref.~\cite{CMSWR}, the most restrictive requiring the invariant mass of the electrons be greater then $ 200 \mbox{GeV} $.

The single gluino production dominates the high mass peak, while the disquark channel contributes broadly to the bins between $ 1 - 2 \mbox{ TeV} $. The broad feature is a consequence of a many-body structure of the decay which, and as pointed out in~\cite{Dobrescu2014}, is useful to evade bounds by the CMS leptoquark search without introducing multiple decay channels. We emphasize that in our model we satisfy both properties of the signal. Firstly, no signal is found in corresponding muon channels as only the electron doublet mixes with the other neutralinos and charginos in this framework. Secondly, the events are dominated by opposite-sign electrons. This is guaranteed by the imposed R symmetry for which an electron and positron have opposite charges. 

% Lastly, we point out that we reduce the second small anomaly in the $ W _R $ search at around $ 1.1 \mbox{TeV} $ through the disquark channel. We did not attempt to fit the masses to this smaller excess since it occurs in a high background region, but nevertheless we take it as a positive sign or the model. 

Next we reproduce the leptoquark (LQ) searches in this framework. In the LQ search a sequence of more stringent cuts are applied, optimized for different mass leptoquarks. In the $ eejj $ channel, the main discriminating variables are $ S _T $ (the scalar sum of $ p _T $ of two leading electrons and jets), $ m _{ee} $  (invariant mass of the two electrons), and $ m _{ ej} ^{ min} $ (the minimum of the electron-jet invariant mass of the four possible combinations for $ eejj $). In the $ e \nu jj $ channel, the main discriminating variables are $ S _T $, $ E ^{ miss} _T $, and $ m _{ ej} $. % , and $ m _{ T , e \nu } $ 
% Generically the null  are the high mass range in these searches are in conflict with the $ W _R $ search results. 
Typically models that predict large $ m _{ eejj} $ (in order to explain the $ W _R $ excess) will also produce large $ S _T $ (and $ m _{ ej} ^{ min} $ unless they arise from a very light LQ). In general, this leads to expected excess in the heavy LQ mass cut range. Thus it is important to check the predictions of any model attempting to explain the CMS excesses in these high mass bins. % However, this tension is alleviated in many jet final states. 
% While this has not been confirmed in previous explanations of the excess we believe this is important for consistency of any explanation of the excess..

\begin{figure}
\centering
\begin{subfigure}[l]{0.48\textwidth}
\includegraphics[width=\textwidth]{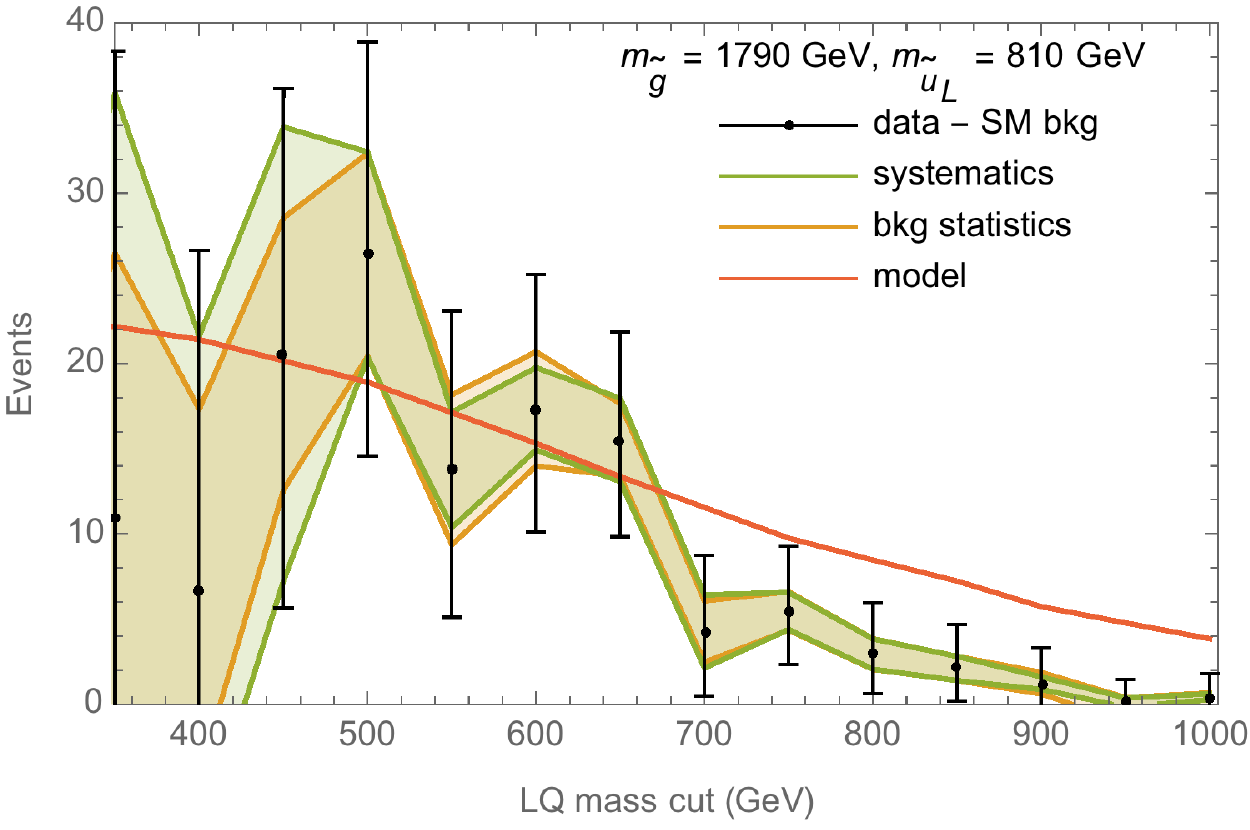}
\caption{$ e e j j $ search}
\label{fig:LQeejj}
\end{subfigure}
\begin{subfigure}[r]{0.48\textwidth}
\includegraphics[width=\textwidth]{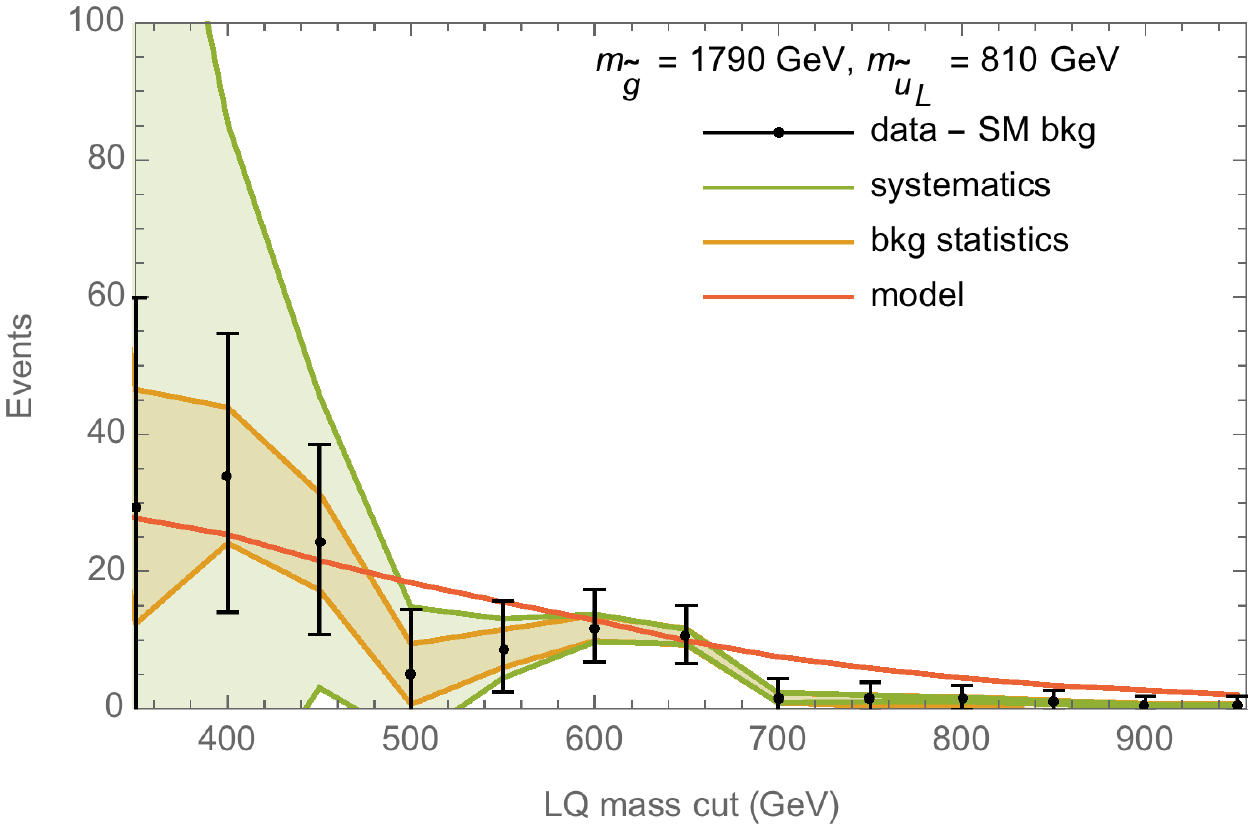}
\caption{$ e \nu j j $ search}
\label{fig:LQevjj}
\end{subfigure}
\caption{Bin-by-bin background-subtracted events for the LQ searches. Each bin count is a subset of the previous bin and hence the bins are highly correlated. The model shows some tension with the data at high LQ mass cuts.} % Model prediction as a function of LQ mass cut in the $ eejj $ and $ e \nu jj $ searches. This bins are highly correlated as each bin contains the events of the previous bin.
\label{fig:tab1}
\end{figure}

The corresponding cuts for each LQ mass can be found in~\cite{CMSLQ} (see tables $2$ and $3$). Here we plot the difference between the data and the SM background as a function of LQ mass cut. The results are shown in Fig.~\ref{fig:tab1}.
Each bin is a fraction of the events in the lower LQ mass cut bin and thus the bins are highly correlated. We see moderate agreement of our signal with the observed counts. We are able to explain the excess in the $ \sim 650 \mbox{GeV} $ region, but see small excess in the higher mass cuts for $ eejj $. The excess in the high mass range is a general characteristic of trying to explain both the $ W _R $ and $ LQ $ searches. Note that the excess is $ {\cal O} ( 5 ) $ events instead of $ {\cal O} ( 10 ) $ which were found in the $ W _R $ search. This is a consequence of the large number of jets increasing the effectiveness of the $ S _T $ cut.

\begin{figure}
\centering
\begin{subfigure}[l]{0.48\textwidth}
\includegraphics[height=4cm]{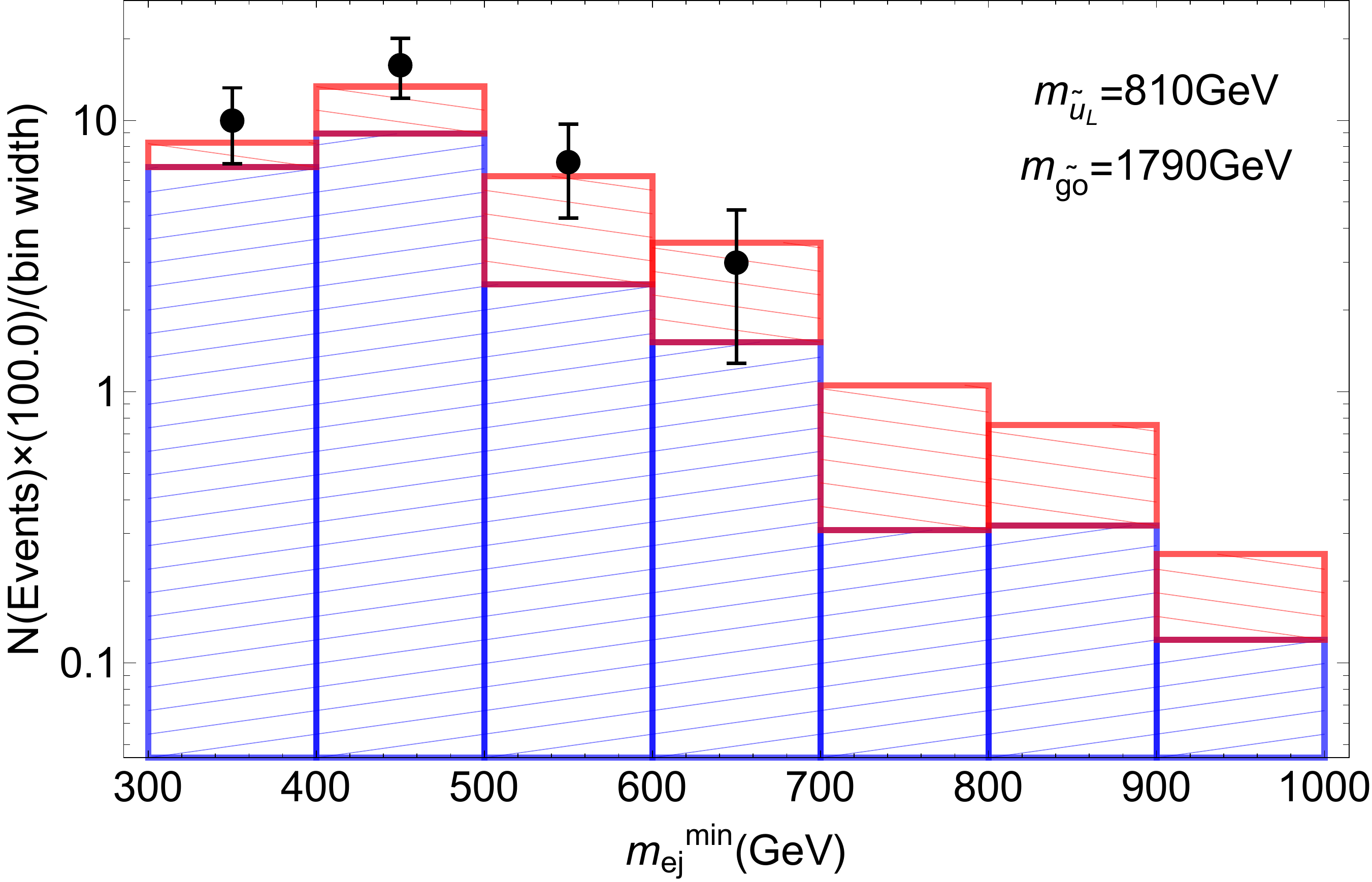}
\caption{$ eejj $ search (figure 5 in~\cite{CMSLQ})}
\label{fig:LQeejj}
\end{subfigure}
\begin{subfigure}[r]{0.48\textwidth}
\includegraphics[height=4cm]{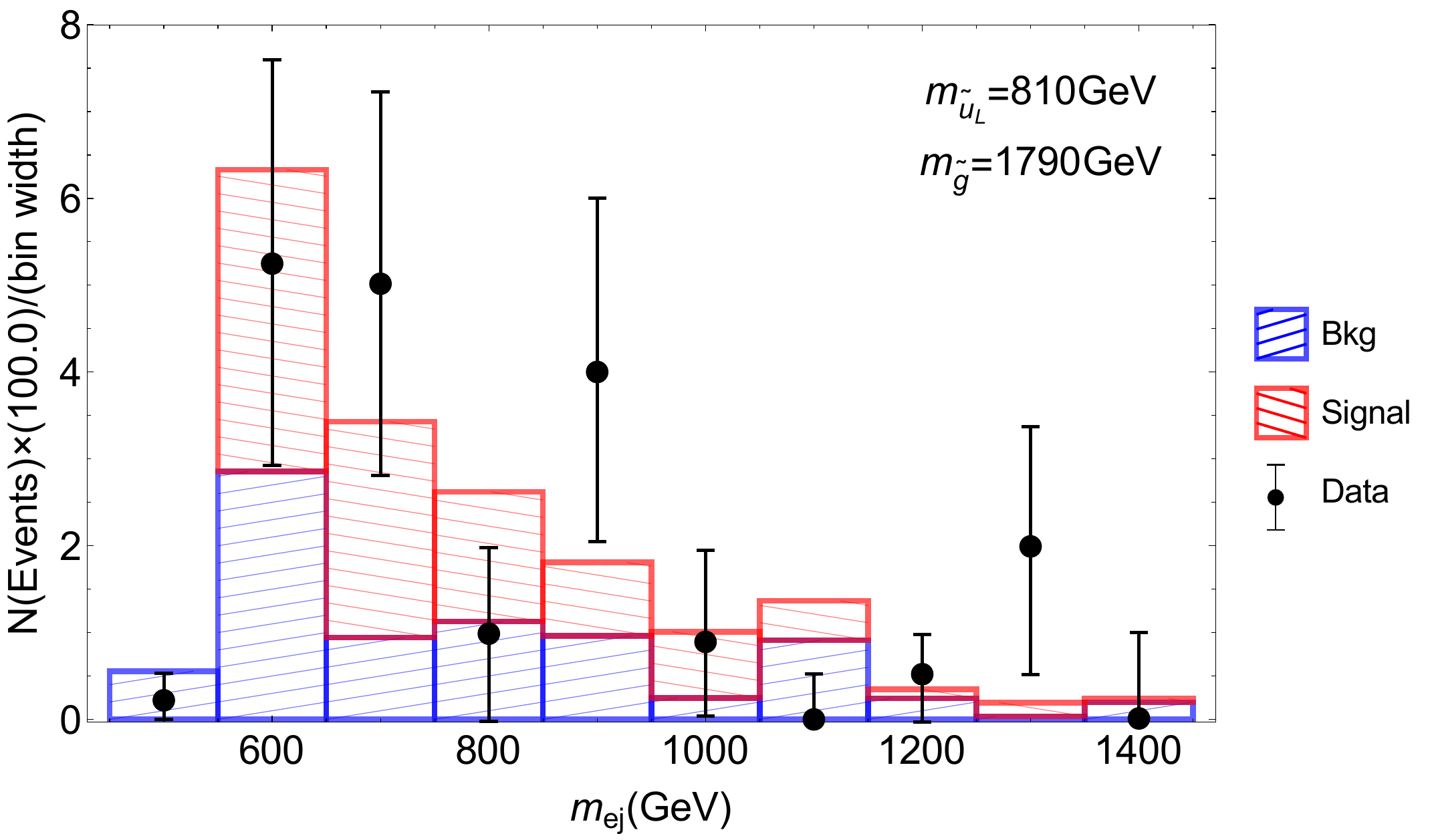}
\caption{$ e \nu jj $ search (figure 8 in~\cite{CMSLQ})}
\label{fig:LQevjj}
\end{subfigure}
\caption{The CMS leptoquark search plots.}
\label{fig:LQ}
\end{figure}

To further check the kinematic properties of the model we compare our $ m _{ ej} ^{ min} $ and $ m _{ ej} $ distributions at the $ 650 \mbox{ GeV} $ mass cut point. The results for both searches are shown in figure \ref{fig:LQ}.
 % \qn{In coloron model Josh didn't show zero points. Should I remove them?}
In both the $ eejj $ and $ e \nu jj $ channel we see good agreement between the model and experiment. The broad feature of the plots is again a consequence of the many jet signal and is necessary to get the right kinematic spread in the LQ invariant mass distributions. %The broadness in the signal is again a consequence of the multibody decay. 
% We note that since we chose the squark to be first generation, we don't need to worry about the final state being charactorized by b-tags, which would disagree with the CMS results. 

% \begin{figure}[!tb]
%   \centering
%   \begin{subfigure}[l]{0.48\textwidth}
% \includegraphics[width=\textwidth]{Hsnu_bins_tab4.pdf}
% \caption{$ eejj $ search}
% \label{fig:bins-eejj}
% \end{subfigure}
% \begin{subfigure}[r]{0.48\textwidth}
% \includegraphics[width=\textwidth]{Hsnu_bins_tab5.pdf}
% \caption{$ e \nu jj $ search}
% \label{fig:binsevjj}
% \end{subfigure}
%   \caption{Bin-by-bin background-subtracted events for the LQ searches.}
%   \label{fig:bin-results}
% \end{figure}

This framework has two characteristic features - many electrons in the final state and many jets. Due to their limited background, we expect the most stringent bounds on our model arise from multilepton searches~\cite{CMSML,ATLASML}. The model produces more than $2$ leptons if each squark decays into an electron and additional leptons arise from vector boson decays. We now roughly estimate the number of expected events in the multilepton searches. The NLO cross section for squark-squark and squark-gluino production at our mass point is $ 5 .7 \mbox{ fb} $. At $ {\cal L} \sim 20 \mbox{ fb}^{-1} $ this corresponds to about 115 events. The probability of both squarks producing electrons (as opposed to neutrinos) is about $ 1/ 4 $. Furthermore, the probability of at least one of the vector bosons decaying leptonically is between $11 $ and $ 40\% $ depending on whether there is a $ WW , WZ , $ or $ ZZ $ is in the final state. This suggests $ 5-10 $ events with $ 3 $ or more leptons. However, these events don't contain any genuine $ E _T ^{ miss} $ or $b$-tagged jets, both of which are powerful discrimating variables in such searches. This makes the signal hard to detect, even in a multilepton search. Thus we conclude the model is safe from current multilepton bounds, though we expect sensitivity with more data at higher energies.

% Such models are expected to have constraints from multilepton searches~\cite{CMSML,ATLASML} as well as many jets + missing transverse energy (MET) searches~\cite{CMSMET,ATLASMET}.In this section we study these constraints and reproduce the bounds with our assumed spectra.

% For the multilepton search we focus on~\cite{CMSML}. We reproduced the event tables and found less than a single event per bin {\color{red} Check this is true for all tables.}. This is reassuring and suggests that the model has not been ruled out by this search. However, we point out that we do expect to see an excess at future LHC runs, in particular in the {\color{red} ...}.

% Next we studied the bounds from many jets + MET searches, with a focus on\cite{CMSMET}. As for the multilepton search we do not find a large excess in any one of the many mutually exclusive event tables. However, we again expect an excess in future runs. In particular, we expect to see events with large missing energy and many jets ($ N _{ jets} \ge 8 $), where we the SM background is expected to be small.

%%% Local Variables:
%%% mode: latex
%%% TeX-master: "late"
%%% End:

\section{Discussion and Conclusions}
%%% Local Variables: 
%%% mode: latex
%%% TeX-master: t
%%% End: 
In this paper, we have explored the phenomenology of a class of SUSY models in which the Higgs is a sneutrino. Such models could account
for excesses seen in the CMS experiment, while accounting for the observed kinematics and flavor structure in a natural way.

As with most SUSY models, several correlated observables are expected.  While the detailed spectrum and branching fractions are model-dependent, these models have a few generic predictions.  Most reliably, there should be correlated excesses in multi-lepton searches.  Since the decay of hadronic sparticles necessarily proceeds via electroweakinos, the decays will generally feature leptons, possibly in large numbers and with a preference for electrons.  These excesses would come with some missing energy from neutrinos, but decays without neutrinos are certainly possible.  The lepton structure of these excesses would again be striking, featuring more electrons than muons or taus.  The scales of $\lesssim 1600~{\rm GeV}$ from $\tilde{q}\tilde{q}^*$, $\lesssim 2400~{\rm GeV}$ from $\tilde{q}\tilde{g}$, and $\lesssim 3600~{\rm GeV}$ from $\tilde{g}\tilde{g}$ would also feature in the total invariant (transverse) mass distribution.

The remaining signals are highly dependent on the more weakly coupled
or heavier elements of the spectrum.  The constraints on sleptons and
electroweakinos remain weak after Run 1 of the LHC, but 
searches for signatures of new electroweak states are a vital part of
Run 2 that can only be fully exploited at high luminosity.  Such
particles with mass $\mathcal{O}(100~{\rm GeV})$ could be in the
spectrum and would decay primarily to elecrtoweak bosons, electrons,
and neutrinos.

The first run of the LHC has seen a remarkable confirmation of the SM
with few  searches finding excesses beyond the $2\sigma$ level.  On
the other hand, several searches that have seen excesses indicate
similar final states with electrons and jets, as well as large energy
scales of $\sim 650~{\rm GeV}$ and $\sim 2~{\rm TeV}$.  If such
excesses are the first hints of a new state beyond the SM, then Run 2
will bring striking and nearly immediate discoveries, as the
sensitivity to physics at $\sim 2~{\rm TeV}$ is vastly superor to that
in the first run. % We therefore turn with great anticipation to the
%first physics results of the Run 2 on channels with electrons and jets.

\acknowledgments{% We thank ... for useful discussions. 
  We thank Yuval Grossman for helpful discussions
  during the completion of this work.
  SLAC is operated by Stanford University for the US Department of
  Energy under contract DE-AC02-76SF00515.  The work of JAD is supported in part by NSERC Grant PGSD3-438393-2013. The work of JAD and WN is
supported is part by the U.S. National Science Foundation through grant PHY-0757868.
}

\bibliographystyle{JHEP}
\bibliography{Draft}

\end{document}